\documentstyle[11pt,aaspp4,nick]{article}
\begin{document}

%\pagestyle{myheadings}
%\markright{DRAFT: \today\hfill}

\title{Local Group Dwarf Galaxies and the Star Formation Law at High Redshift}
\author{Nickolay Y.\ Gnedin}
\affil{CASA, University of Colorado, Boulder, CO 80309;
e-mail: \it gnedin@casa.colorado.edu}

%$$\framebox{$\displaystyle\phantom{\prod}$DRAFT: \today}$$

\load{\scriptsize}{\sc}

\def\A{{\cal A}}
\def\B{{\cal B}}
\def\ion#1#2{\rm #1\,\sc #2}
\def\HI{{\ion{H}{i}}}
\def\HII{{\ion{H}{ii}}}
\def\GI{{\ion{He}{i}}}
\def\GII{{\ion{He}{ii}}}
\def\GIII{{\ion{He}{iii}}}
\def\MH{{{\rm H}_2}}
\def\Hp{{{\rm H}_2^+}}
\def\Hm{{{\rm H}^-}}

\def\dim#1{\mbox{\,#1}}

\def\figdir{.}
\def\placefig#1{#1}

\begin{abstract}
I show how the existing observational data on Local Group dwarf galaxies 
can be used to
estimate the average star formation law during the first $3\dim{Gyr}$ of
the history of the universe.
I find that the observational data are consistent with the 
orthodox Schmidt law with a star formation efficiency of 
about 4 percent if the star formation is continuous (during the
first $3\dim{Gyr}$). The efficiency is
proportionally higher if most of the gas in the dwarfs
was consumed (and never replenished) in a short time interval
well before the universe turned $3\dim{Gyr}$.
\end{abstract}


\section{Introduction}

Direct observational measurements of the star formation rate in the very first
objects in the universe is exceptionally difficult, and even with the recent 
observational and technological 
advances, is well beyond reach. Nevertheless, it is possible to reconstruct
the general features of the star formation at high redshift in much the same
way as paleontologists reconstruct the prehistoric life from fossils.

In the 
currently widely
accepted hierarchical clustering paradigm the characteristic mass scale of
cosmological objects increases with time: low mass objects form first, and
more massive objects form later. Thus,
a considerable 
fraction of the dwarf galaxies may be very old, and may
represent the fossils of 
the very first star formation in the universe. Although these dwarf galaxies
are small and faint, and thus hard to observe, a vast 
amount of detailed information exists on the Local Group dwarf galaxies,
as summarized in several recent reviews (Grebel 1998; Mateo 1998;
van den Bergh 1999). The first two reviews are particularly useful for my
purpose, as they contain the star formation histories of a large fraction of
all Local Group dwarfs. A careful investigation of the available data
reveals a striking feature common to the vast majority (more than 90\%)
of all dwarf galaxies
with known star formation histories: a sharp decline in the star formation
rate about $10\dim{Gyr}$ ago.

There are at least five possible explanations for such a feature in the
star formation history. First,
it may be attributed to the tidal effects from the Galaxy or Andromeda.
However, in the hierarchically clustering universe neither the Galaxy nor
Andromeda existed at that time as entities, but rather as an interacting
collection of proto-galactic clumps (see, for example, Contardo, Steinmetz,
\& Fritze-von Alvensleben 1998), and thus they were not able to have a large
impact on the star formation histories of {\it almost all\/} dwarf galaxies.

A second possible explanation is the effect of supernovae, which could have
expelled the gas from the low potential wells and quenched the star formation
rate. However, it is hard to believe that the supernova activity was so
well synchronized in 90\% of all dwarf galaxies as to expel the gas
almost simultaneously (this is elaborated further in the Conclusions).

Since the drops in the star formation rates in dwarf galaxies are essentially 
simultaneous, it is plausible that this was triggered by a
process that affected the whole universe. One such process
is reionization. It is well
established that photoheating of the intergalactic medium during 
cosmological reionization sharply reduces the gas fractions in the low
mass objects (Thoul \& Weinberg 1996; Quinn, Katz, \&
Efstathiou 1996; Weinberg, Hernquist, \& Katz 1997; Navarro \& Steinmetz
1997; Gnedin 2000). This, in turn,
will lead to the drop in the star formation rates in dwarf galaxies,
including the Local Group dwarfs. This proposal however encounters a
difficulty because an age of $10\dim{Gyr}$ corresponds to
about $z=2$ (in a $13\dim{Gyr}$ old universe with $h\Omega_0^{1/2}=0.4$),
which is too late for cosmological reionization. However, 
if the drop in the star formation rate in dwarf
galaxies occurred some twelve billion years ago,
the reionization explanation could work. This requires that 
the measured stellar
ages are off by about $2\dim{Gyr}$, but the question of whether this is
plausible or not is outside the scope of this paper.

The fourth explanation is based on the observational evidence (which is
admittedly not completely compelling) that the
universe experienced a second epoch of reheating at about $z=3$ 
(Ricotti, Gnedin, \& Shull 2000; Schaye et al.\ 2000). This second
reheating (plausibly attributed to the reionization of helium) has
a similar effect on dwarf galaxies as hydrogen reionization at
$z\sim7-10$, except that it takes place somewhere between $z=2$ and $z=3$.

It is also possible that photoionization associated with the burst of star
formation that made the bulges of the Galaxy and Andromeda 
(``local reionization'') inhibited the
star formation in the surrounding dwarfs (van den Bergh 1994). 
While possible in principle, it is not clear whether this explanation
is plausible within the hierarchical clustering paradigm, or whether
it agrees with the observed metallicity distribution of the Galactic
bulge (McWilliam \& Rich 1994).

Whatever the explanation of the observed simultaneous drop in the star
formation rate in 90\% of Local Group dwarfs is, it is not essential
to the measurement described in this paper. The only thing that matters
is that there exists a subset of the Local Group dwarfs that had
little star formation in the last $10\dim{Gyr}$, so that their stellar
content provides a fossil record of the early universe.

\section{Results}

Because the star formation histories of the Local Group dwarfs are quite
diverse, only a subset of all galaxies can be used, namely those
that have a large fraction
of their stars formed more than $10\dim{Gyr}$ ago. This limits the
available sample to seventeen out of thirty two shown in Fig.\ 8 of
Mateo (1998) and Fig.\ 5-6 of Grebel (1998).
Out of a total of seventeen, seven (Fornax, Sextans B, NGC 185, Phoenix, GR 8, 
Draco, and  Tucana) either have some of the
essential data missing or Mateo (1998), Grebel (1998) and
Hernandez, Gilmore, \& Valls-Gabaud (2000)
strongly disagree
on their star formation histories,  
so only the ten listed in Table \ref{tabone} are used in the present
analysis. The other seven can be added to this analysis when the missing
or disgreed upon data become available.

\def\tableone{
\begin{deluxetable}{ccccc}
\tablecaption{Local Group Dwarfs Used in this Analysis\label{tabone}}
\tablehead{
\colhead{Galaxy} & 
\colhead{$R_{10}$\tablenotemark{a}} & 
\colhead{$\log(L_V/(I_cr_c^3))$\tablenotemark{b}} & 
\colhead{$\log(I_c\dim{pc}^{3}/L_{\sun})$\tablenotemark{b,c}} & 
\colhead{$f_{>10}$\tablenotemark{d}} }
\startdata
NGC 3109   & 0.2-0.5\tablenotemark{e} 
                 & $1.55\pm0.23$ & $-1.75\pm0.20(0.33)$ & 0.7-0.5
						\tablenotemark{e} \\
NGC 205    & 0.2 & $1.69\pm0.27$ & $-0.37\pm0.20(0.25)$ & 0.8 \\
NGC 147    & 0.3 & $1.84\pm0.15$ & $-0.41\pm0.12(0.17)$ & 0.7 \\
Ursa Minor & 0-0.2\tablenotemark{e}   
                 & $0.78\pm0.26$ & $-2.22\pm0.21(0.23)$ & 0.9 \\
Sextans    & 0.1 & $0.82\pm0.24$ & $-2.70\pm0.21(0.28)$ & 0.8-0.7
						\tablenotemark{e} \\
Antlia     & 0-0.2\tablenotemark{e}   
                 & $0.95\pm0.12$ & $-1.80\pm0.10(0.12)$ & 0.9 \\
Sculptor   & 0.1 & $1.47\pm0.20$ & $-1.26\pm0.17(0.19)$ & 0.9 \\
And I      & 0-0.1\tablenotemark{e}            
                 & $1.00\pm0.12$ & $-2.05\pm0.10(0.11)$ & 0.9 \\
And II     & 0.2 & $1.21\pm0.12$ & $-1.77\pm0.23(0.25)$ & 0.9 \\
And III    & 0-0.1\tablenotemark{e}   
                 & $1.03\pm0.14$ & $-1.75\pm0.10(0.11)$ & 0.9 \\
\enddata
\tablenotetext{a}{The ratio of the star formation rate just after the drop
at $10\pm1\dim{Gyr}$ ago to the star formation rate just before the drop.
Estimated from Fig.\ 8 of Mateo (1998) or Fig.\ 5-6 of
Grebel (1998). This column is shown
for illustration only, and is not used in the analysis.}
\tablenotetext{b}{The data are taken from Tables 2-4 of Mateo (1998).}
\tablenotetext{c}{The errors in parenthesis
are corrected by a factor of %1/f_{>10}$, 
and outside of parenthesis are uncorrected.}
\tablenotetext{d}{The fraction of all stars formed more than $10\dim{Gyr}$ ago.
Estimated from Fig.\ 8 of Mateo (1998) or Fig.\ 5-6 of Grebel (1998).}
\tablenotetext{e}{The first number is from Mateo (1998) and the second one is 
from Grebel (1998); for $f_{>10}$ the arithmetic mean is used in the analysis.}
\end{deluxetable}
}
\placefig{\tableone}
According to the Schmidt law, on sufficiently large scales the star formation
rate can be parametrized as a power-law of the gas density,
\begin{equation}
	{d\rho_*\over dt} \propto \rho_g^n,
	\label{rhon}
\end{equation}
where $\rho_*$ is the mass density of stars, and $\rho_g$ is the mass density
of gas. 
Integrating equation (\ref{rhon}) over the total volume of a galaxy
and time,
and assuming that the mass-to-light ratio is independent of radius, I can
obtain the following equation connecting the total luminosity of a galaxy at
some moment $t_f$ with
the total baryon core density $\rho_{b,c}$ at the same time,
\begin{equation}
	L_* \propto \Delta t_{\rm eff}\rho_{b,c}^n r_c^3 4\pi \int_0^\infty 
	\left(\rho_b\over \rho_{b,c}\right)^n 
	\left(r\over r_c\right)^2 {dr\over r_c},
	\label{lsrg}
\end{equation}
where $L_*$ is the luminosity of a dwarf galaxy, $r_c$ is the core radius,
and $\Delta t_{\rm eff}$ is the effective
time interval, $\Delta t_{\rm eff} \equiv \int_0^{t_f} \int \rho_g^n dV\, dt
/\int \rho_b^n(t_f) dV$,
which incorporates all the details of the increase (due to infall) and
decrease (due to consumption by star formation) of the gas.

It is important that I use the core baryon density rather than core gas 
density. The gas in the center can be almost entirely consumed, whereas
the baryon density will likely stay the same (unless there is diffusion
of stars from the center outward).

If all dwarf galaxies are
structurally similar (which means that the integral in eq.\ [\ref{lsrg}] is the
same for all dwarfs), and in the core almost all the gas is transformed into
stars ($\rho_{b,c}\approx\rho_{*,c}$, this assumption is addressed below), 
then equation (\ref{lsrg}) reduces
to the power-law dependence between the total luminosity of a galaxy
and its central luminosity density $I_c$,
\begin{equation}
	L_*/r_c^3 \propto I_c^n
	\label{lsic}
\end{equation}
at $t=t_f$. However, this equation also holds at the present time if the
star formation at later times ($t>t_f$) is negligible or can be corrected
for.

It is important to point out here, that it is the luminosity density
(i.e.\ the stellar mass density) and not the total mass density that
matters. The dark matter density in the Local Group dwarfs varies
enormously and is not correlated to the stellar density, as can be easily seen
from the total mass-to-light ratios (Table 4 of Mateo 1998). Since the
galaxies included in Table \ref{tabone} formed most of their stars more
than $10\dim{Gyr}$ ago, it can be presumed that the observed core luminosity
densities of the dwarfs are representative of the core luminosity densities
at $t_f\approx 3\dim{Gyr}$. This cannot be said about the dark matter, however.
In the last $5\dim{Gyr}$ or so the Local Group dwarfs inhabited a region
with the mean density about hundred times higher than the mean density of the
universe, and therefore they had enough opportunities to accrete large
quantities of dark matter. Thus, their dark matter contents is
representative of their whole evolutionary history, and not of only
the first $3\dim{Gyr}$ of their life.

The data from Table \ref{tabone} can be used to investigate the 
relationship (\ref{lsic})
in the Local Group dwarfs. It is natural to take $t_f$ as the
time $10\dim{Gyr}$ ago, corresponding to the sharp drop in the star 
formation rate in all galaxies in Table \ref{tabone}. For a $13\dim{Gyr}$
old universe this would imply $t_f=3\dim{Gyr}$.

\def\capLI{
The corrected V-band luminosity per cube of the core radius versus
the central luminosity density for the Local Group dwarfs from Table
\protect{\ref{tabone}}. The dotted line shows the power-law fit with 
the slope fixed to $3/2$.}
\placefig{
\begin{figure}
%\epsscale{0.70}
\insertfigure{\figdir/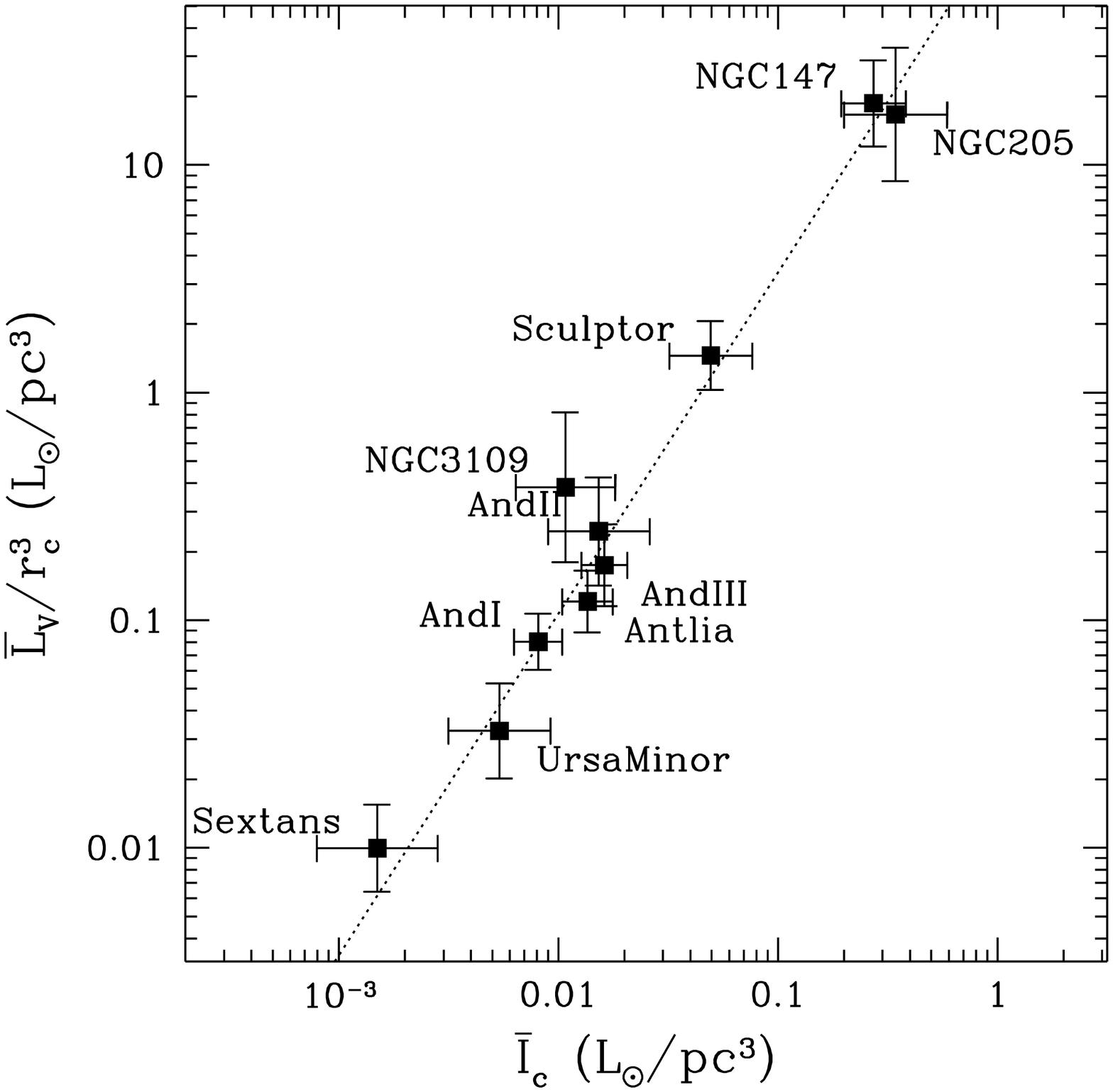}
\caption{\label{figLI}\capLI}
\end{figure}
}
However, even if the galaxies in Table \ref{tabone} formed most of their stars 
more than $10\dim{Gyr}$ ago, they still had additional star formation at 
later times. These ``recent'' stars need to be excluded from the analysis.
It is straightforward to estimate the fraction $f_{>10}$ of all stars
formed more than $10\dim{Gyr}$ ago, from Fig.\ 8 of Mateo (1998) and
Fig.\ 5-6 of Grebel (1998). The quantity $f_{>10}$ is listed in the last
column of Table \ref{tabone}. Thus, instead of $L_*$ and $I_c$ from
equation \ref{lsic}, I use $\bar{L}_*=L_*f_{>10}$ and
$\bar{I}_c=I_cf_{>10}$ instead. Figure \ref{figLI} shows the relationship
(\ref{lsic}) for the data from Table 1 together with a power-law fit with 
the slope $n=3/2$. 

However, Fig.\ \ref{figLI} is a bad way to represent the data because the 
errors along the two axes are not independent - both quantities include the
large errors due to uncertainties in the distances and the core radii. It is 
better to represent the relationship (\ref{lsic}) as
\begin{equation}
	y \equiv {L_*\over I_cr_c^3} \propto I_c^{n-1}.
	\label{lsicbest}
\end{equation}
In this form the left hand side is dimensionless, and thus independent of the
distance, whereas the errors on $I_c$ are dominated by the error in the
distance, 
because the central luminosity density is inversely
proportional to the distance cubed.
In this form the errors on $y$ and $I_c$ are not strongly correlated, and I
can measure the statistical significance of the correlation between the
two quantities in equation (\ref{lsicbest}).

Given the the errors
on the observed quantities: the relative photometric error $\delta V$ in
the total luminosity, 
the relative error in the measured angular core radius $\delta r_c$, the 
relative distance error $\delta D$, and
the relative photometric error in 
the observed central surface brightness $\delta \Sigma_c$, the errors
on $y$ and $I_c$ are calculated as 
$\delta y^2 = \delta V^2 + 4\delta r_c^2 + \delta \Sigma_c^2$, and
$\delta I_c^2 = \delta\Sigma_c^2 + \delta r_c^2 + 9\delta D^2$.

However, my correction to the total luminosity and the central luminosity 
density is, admittedly, 
uncertain, because the star formation histories are not known
too accurately, and
thus the quantity $f_{>10}$ is somewhat uncertain as well. To 
account for this uncertainty, I increase the errors on 
$\bar{I}_c$ a factor of $1/f_{>10}$, 
in which case 
the contribution of dwarfs with a
large correction to the total $\chi^2$ 
is reduced by a factor of $f_{>10}^2$. 

The best 
power-law fit to the data listed in Table \ref{tabone}
is given by the following formula:
\begin{equation}
	{\bar{L}_V\over r_c^3} = 10^{1.19\pm0.06} \bar{I}_c 
	\left(\bar{I}_c\over 0.02 L_{\sun}/\dim{pc}^3\right)^{0.54\pm0.11}.
	\label{bestfit}
\end{equation}
In this form the errors on the amplitude and the slope are uncorrelated.
If, instead, I use uncorrected errors, 
the reduction of the fit error is minor:
the power-law index is $0.52\pm0.10$, and the logarithm of the amplitude
is $1.20\pm0.06$,
which demonstrates that my correction procedure is
not too important. I can also fit the original data, without correcting the
total luminosity and the central luminosity density by a factor $f_{>10}$.
In the latter case I obtain an equally good fit:
the power-law index is $0.51\pm0.10$, and the logarithm of the
amplitude is $1.16\pm0.06$. The reason 
why the original data work equally well is that the fit is dominated by 
the data
points with the smallest error-bars, and those also have the smallest
correction. Whether this is a mere coincidence, or has a physical meaning,
can only be speculated at the moment.

All three fits are highly significant statistically, which indicates that the
error bars are not underestimated.

The result (\ref{bestfit}) is non-trivial. It shows that the {\it average\/}
luminosity density of dwarf galaxies was proportional to the $3/2$ power
of their {\it central\/} luminosity density in the first $3\dim{Gyr}$.
The central luminosity density, as I argue below, is representative of
their original gas density, since the star formation at the center is
efficient enough to be able to convert all their gas into stars. The
average luminosity density (i.e.\ the total luminosity) of the dwarfs
is however dominated by the outer regions (as can be seen from
Table \ref{tabone} since $y\gg1$).
In the outer regions the star formation
is inefficient,
and thus the total luminosity is proportional to the gas density (and thus
the central luminosity density) to the $3/2$ power (or about that), and
not to the central luminosity density to the first power, which would be
the case if the star formation was efficient everywhere and all the gas
in the dwarfs was converted into stars.

It is remarkable that the observed slope is so close to the orthodox Schmidt
law power-law index of $3/2$. Therefore, it is worthwhile to investigate this
special case in more detail. The star formation rate in this case can
be rewritten in the following form, fixing the coefficient of proportionality
in the Schmidt law:
\begin{equation}
	{d\rho_*\over dt} = \epsilon {\rho_g\over t_{\rm dyn}},
	\label{rhoeos}
\end{equation}
where a dimensionless quantity $\epsilon$ measures the efficiency of star
formation, and the dynamical time $t_{\rm dyn}$ is defined as in 
Binney \& Tremaine (1987),
$t_{\rm dyn} \equiv (3\pi/[16 G \rho_g])^{1/2}$.
%\begin{equation}
%	t_{\rm dyn} \equiv \sqrt{3\pi\over 16 G \rho_g}.
%	\label{tdyn}
%\end{equation}
Note, that I {\it define\/} the dynamical time as depending 
on the gas density, and not on the total
density. If one desires to define the dynamical time in
terms of the total density, $\epsilon$ simply needs to be decreased by a factor
of $(\Omega_0/\Omega_b)^{1/2}$.

Integrating equation (\ref{rhoeos}) over time and space, and again
invoking $\Delta t_{\rm eff}$ to hide the consumption or infall of
gas, I obtain the
following relationship between the total luminosity of a galaxy and the
core baryon density:
\begin{equation}
	L_* =  
	3.37\times10^{-4} \dim{cm}^{3/2}\dim{g}^{-1/2}\dim{s}^{-1}
	4\pi \Upsilon^{-1} \Delta t_{\rm eff} \epsilon
	\rho_{b,c}^{3/2} r_c^3 C_{3/2},
	\label{lsict}
\end{equation}
where $\Upsilon$ is the mass-to-light ratio,
and $C_{n} \equiv \int_0^\infty 	
	\left(\rho_b/\rho_{b,c}\right)^n
	r^2\,dr/r_c^3.$
For the King model with the concentration parameter 2.5,
$C_{3/2}=0.6$. Since, as I argue below, the central luminosity density
in the dwarfs is representative of the central baryon density (i.e.\
almost all of the gas is consumed at the center), the King profile
is a good approximation to the total baryon density profile near the
center. The Navarro, Frenk, \& White (1997) density profile, motivated by 
cosmological simulations, also gives $C_{3/2}=0.6$, 
which indicates that this number is not very sensitive to the details of
the density profile. 

%Cosmological simulations offer another way of estimating the factor 
%$C_{3/2}$. As is well known, the dark matter density profile of a
%virialized object in a hierarchical clustering universe is given (at least
%outside the core radius, which is defined as the radius at which the
%local slope of the density profile is equal to -2) by
%the Navarro, Frenk, and White (1997) fit,
%$$
%	\rho_b(r) = {\rho_{b,c}\over r/r_c(1+r/r_c)^2}.
%$$
%Assuming this form of the baryon density profile for $r>r_c$, where the gas 
%is expected to follow the dark matter, I can add the
%following profile for the core,
%$$
%	\rho_b(r) = {\rho_{b,c} \over (1+r^2/r_c^2)^2}
%$$
%for $r<r_c$. Thus obtained profile is continuous and has a continuous first
%derivate at $r=r_c$. For this profile the factor $C_{3/2}$ is also equal
%to 0.6, which indicates that it is not very sensitive to the details of
%the density profile. 

Finally, I can relate the core baryon density to the core luminosity density
by $\rho_{b,c} = \Upsilon I_c/f_*$,
where $f_*$ is the fraction of baryons turned into stars at the center.

Comparing equation (\ref{bestfit}) taken with the power-law index $3/2$
with equation (\ref{lsict}), I obtain the expression for the value of the
star formation efficiency $\epsilon$:
\begin{equation}
	\epsilon = 0.040\times10^{\pm0.06} \Delta t_2^{-1} \Upsilon_4^{-1/2}
	f_*^{3/2},
	\label{epsfstar}
\end{equation}
where $\Delta t_2=\Delta t_{\rm eff}/2\dim{Gyr}$, and 
$\Upsilon_4=\Upsilon/(4M_{\sun}/L_{\sun})$.

Using equations 
(\ref{rhoeos}) and (\ref{epsfstar}),
I can obtain the following expression
for the star formation time $\tau_{*,c}\equiv \rho_{*,c}/(d\rho_{*,c}/dt)$
at the center of a dwarf galaxy:
$$
	\tau_{*,c} = 1.5\times10^9\dim{Gyr}\Delta t_2 I_2^{-1/2}
	(1-f_*)^{-3/2},
$$
where $I_2\equiv I_c/(0.01L_{\sun}/\dim{pc}^3)$. Thus, for all dwarfs
with $I_c\ga2\times10^{-2}L_{\sun}/\dim{pc}^3$ (which are all galaxies listed
in Table \ref{tabone} with the exception of the Sextans and Ursa Minor
dwarf spheroidals)
$\tau_{*,c} \la t_f$ until most of the gas is consumed and
$f_*$ becomes close to unity (an average $1\sigma$ error on $I_c$ corresponds
to about 40\% error on $f_*$, so $f_*=0.6$ and $f_*=1$ cannot be
distinguished on the basis of the current data).
Thus, assuming $f_*=1$ (which is certainly true for the most dense
dwarfs with $I_c\gg2\times10^{-2}L_{\sun}/\dim{pc}^3$), I
obtain the final expression for the average efficiency of star formation
in the first $3\dim{Gyr}$:
\begin{equation}
	\epsilon \Delta t_2= 0.040\times10^{\pm0.06}  
	\Upsilon_4^{-1/2}.
	\label{fineps}
\end{equation}
While $\Upsilon_4$ is unlikely to be different from unity, the last
remaining factor $\Delta t_2$
is less certain. It cannot be larger than 1.5, since 
the gas density cannot exceed the total baryon density.
In particular, $\epsilon > 0.022\Upsilon_4^{-1/2}$ at 3 $\sigma$.
However, $\Delta t_2$ can be much smaller than 1
(giving $\epsilon\gg0.04$) if most of the gas in the dwarfs
was consumed (and never replenished) in a short time interval
well before the universe turned $3\dim{Gyr}$ (it is not clear whether
the existing data rule out such a scenario).

\section{Conclusions}

I use a sample of Local Group dwarf galaxies selected by the criterion that
they form most of their stars more than about $10\dim{Gyr}$ ago to measure
the star formation law in the first $3\dim{Gyr}$ of the life of the universe
(assuming the age of the universe is $13\dim{Gyr}$). I find that the
star formation rate on scales comparable to the sizes of dwarf galaxies
($100-600\dim{pc}$) is a power-law function of the gas density, with the
power-law index $1.54\pm0.11$. If the index is fixed at $n=3/2$, the
star formation rate is given by equations(\ref{rhoeos})
with an efficiency of about 4\% if the gas is consumed continuously.
The efficiency cannot be smaller than 2.2\% but can be higher if the
gas was consumed well before the universe turned $3\dim{Gyr}$.

The efficiency is not a strong
function of scale for scales in the range from about 100 to 600 parsecs
(if it were, the straight power-law would not be a good fit in
Fig.\ \ref{figLI}). 

Admittedly, the time of the drop in the star formation rate in the Local Group
dwarfs is not very well determined, and is, perhaps, uncertain to about one
or two billion years.  Thus, the ``simultaneous'' drop in star formation
rates in the majority of the dwarfs is perhaps not sufficiently
simultaneous to preclude the possibility that it was supernova driven winds
rather than reheating of the universe (or local reionization), 
which was responsible for the observed
drop. However, in this case it would be difficult to explain the observed
correlation, since the central regions of the dwarfs galaxies can be
expected to be affected most by the supernova activity, thus spoiling the
beautiful correlation shown in Fig.\ \ref{figLI}.
Therefore, the present result indicates that
the supernova driven winds play only a modest role in the early evolution
of low-mass galaxies. 

Curiously, equation \ref{lsicbest}) can be used, in principle, as a
novel distance indicator, since only the $x$-axis depends on the distance.
However, it is likely to not be competitive with other indicators given
the amount of information needed to measure $\bar{L}_V$ and $\bar{I}_c$.

The result presented here can be incorrect if the structural form
of the Local Group dwarfs changes monotonically with their luminosity
as to mimic the observed relationship (\ref{bestfit}). I would like to 
emphasize that, while this possibility
cannot be completely excluded, 
there is currently no observational evidence that supports 
it, and this assumption also contradicts the modern theoretical view on
the structure of galaxies.

\acknowledgements

I am grateful to John Bally, Ed Bertschinger, Brad Gibson, Oleg Gnedin, 
Phil Maloney, Mike Shull, John Stocke, and the anonymous referee
for useful comments and discussions.

\placefig{\end{document}}

\clearpage

\newcounter{figurecap}
\setcounter{figurecap}{0}

\begin{center}
\bf Figure Captions
\end{center}

\refstepcounter{figurecap}
Fig.\ \thefigurecap---\label{figLI}\capLI

\clearpage

\tableone


\begin{references}

\reference{BT87}
Binney, J., \& Tremaine, S. 1987, Galactic Dynamics (Princeton: Princeton 
University Press), 37

\reference{CSF98}
Contardo, G., Steinmetz, M., \& Fritze-von Alvensleben, U. 1998, \apj, 507, 497

\reference{G00}
Gnedin, N.\ Y. 2000, \apj, submitted (astro-ph/0002151)

\reference{G98}
Grebel, E. 1998, IAUS, 192, 1

\reference{HGV00}
Hernandez, X., Gilmore, G., \& Valls-Gabaud, D. 2000, \mnras, in press
(astro-ph/0001337)

\reference{M98}
Mateo, M. 1998, \araa, 36, 435

\reference{MR94}
McWilliam, A., \& Rich, R.\ M. 1994, \apjs, 91, 749

\reference{NFW97}
Navarro, J., Frenk, C.\ S., \& White, S.\ D.\ M. 1997, \apj, 490, 493

\reference{NS97}
Navarro, J., \& Steinmetz, M. 1997, \apj, 478, 13

\reference{QKE96}
Quinn, T., Katz, N., \& Efstathiou, G. 1996, \apj, 278, 49P

\reference{RGS99}
Ricotti, M., Gnedin, N.\ Y., \& Shull, J.\ M. 2000, \apj, in press
(astro-ph/9906413)

\reference{Sea00}
Schaye, J., Theuns, T., Rauch, M., Efstathiou, G., \& Sargent, W.\ L.\ W. 2000,
\mnras, submitted (astro-ph/9912432)

\reference{TW96}
Thoul, A.\ A., \& Weinberg, D.\ H. 1996, \apj, 465, 608

\reference{B94}
van den Bergh, S. 1994, \apj, 428, 617

\reference{B99}
van den Bergh, S. 1999, \aa Rev., 9, 273

\reference{WHK97}
Weinberg, D.\ H., Hernquist, L., \& Katz, N. 1997, \apj, 477, 8

\end{references}
\end{document}